# A Multi-Species Model for Hydrogen and Helium Absorbers in Lyman-Alpha Forest Clouds

Yu Zhang, Peter Anninos and Michael L. Norman

*Laboratory for Computational Astrophysics*

*National Center for Supercomputing Applications*

*University of Illinois at Urbana-Champaign*

*405 N. Mathews Ave., Urbana, IL 61801*

– 2 –## ABSTRACT

We have performed a multi-species hydrodynamical simulation of the formation and evolution of Ly$\alpha$ clouds in a flat CDM dominated universe with an external flux of ionizing radiation. We solve the fully coupled non-equilibrium rate equations for the following species: $H$, $H^+$, $H^-$, $H_2$, $H_2^+$, $He$, $He^+$, $He^{++}$, and $e^-$. The statistical properties of the distribution and evolution of both hydrogen and helium absorption lines are extracted and compared to observed data. We find excellent agreement for the following neutral hydrogen data: the distribution of column densities is fit well by a power law with exponent $\beta = 1.55$ with a possible deficiency of lines above column density $10^{15}$ $cm^{-2}$; the integrated distribution matches observed data over a broad range of column densities $10^{13}$ to $10^{17}$ $cm^{-2}$; a Gaussian statistical fit to the Doppler parameter distribution yields a median of 35.6 $km$ $s^{-1}$; the evolution of the number of clouds with column densities larger than $10^{14}$ $cm^{-2}$ follows a power law with exponent $\gamma = 2.22$. Analogous calculations are presented for $HeII$ absorption lines and we find the ratio of Doppler parameters $b_{HeII}/b_{HI} = 0.87$. Our data also suggests that Ly$\alpha$ clouds may belong to two morphologically different groups: small column density clouds which tend to reside in sheets or filamental structures and are very elongated and/or flattened, and the large column density clouds which are typically found at the intersections of these caustic structures and are quasi-spherical.

– 3 –## 1. INTRODUCTION

Quasar absorption lines (Ly$\alpha$ forest, Lyman limit and damped Ly$\alpha$) are becoming increasingly important because of the enormous amount of observational data and because of their uniqueness in probing the early stages of cosmic structure formation and evolution. Also, recent observations of $HeII$ (Jakobsen et al. 1994; Davidsen et al. 1995; Tytler et al. 1995) provide complimentary tools for discriminating between different cosmological models and for diagnosing the physical state in the pregalatic and intergalactic medium. Although tremendous progress has been made in observational studies of the Ly$\alpha$ clouds, progress in the theoretical understanding of the structure and origin of the forest has proceeded more slowly. Recent numerical calculations (Cen et al. 1994; Weinberg et al. 1995) have provided crucial insights into the nature of the clouds, suggesting that the gravitational collapse of small scale structure and photoionization can explain the statistical properties of the absorbers.

We have developed a two-level hierarchical three-dimensional numerical code for cosmology (HERCULES), designed to simulate structure formation in an expanding dark matter dominated universe with Newtonian gravity, multi-fluid hydrodynamics, radiative cooling, non-equilibrium kinetics and external radiation fields (Anninos, Norman & Clarke 1994; Anninos, Zhang & Norman 1995). The following nine species are tracked separately: neutral hydrogen $HI$, ionized hydrogen $HII$, negatively-charged hydrogen $H^-$, hydrogen molecules $H_2$, ionized hydrogen molecules $H_2^+$, neutral helium $HeI$, once-ionized helium $HeII$, twice-ionized helium $HeIII$ and free electrons $e^-$. The non-equilibrium chemical rate equations (27 in all, including the radiation processes) are solved in generality for the abundances of each species. The rate coefficients used in the chemistry model are provided in Abel & Norman (1995). We have also implemented a comprehensive model for the radiative cooling of the gas. Our cooling model includes atomic line excitation, recombination, collisional ionization, free-free transitions, molecular line excitations, and Compton scattering of the cosmic background radiation (CBR) by electrons.

We apply our model to high redshift pre-galactic structure formation and evolution, looking specifically at Ly$\alpha$ forest clouds. Our effort emphasizes the modeling of various microphysical processes in the gas as structures collapse gravitationally within the framework of a standard CDM spectrum of fluctuations. Our model background spacetime is a flat ($\Omega_0 = 1$) dark matter dominated universe with Hubble constant $H_0 = 70\ km\ s^{-1}\ Mpc^{-1}$ and baryonic fraction $\Omega_B = 0.04$, consistent with the latest constraints from big-bang nucleosynthesis (Copi et al. 1994). The baryonic matter is composed of hydrogen and helium in cosmic abundance with a hydrogen mass fraction of 76%. The initial data is the Harrison-Zel'dovich power spectrum modulated with a transfer function appropriate to cold dark matter (CDM) adiabatic fluctuations and normalized to a bias factor of $b = 1.0$. A uniform UV ionizing background is included, with a flux given by $F_\nu \sim (\nu/\nu_{th})^{-\alpha}$, with spectral index $\alpha = 1.5$, and an amplitude at the Lyman limit of $F_\nu = 10^{-21}\ ergs\ s^{-1}\ cm^{-2}\ Hz^{-1}$. We begin the simulations at redshift $z = 50$ and turn on the UV radiation at redshift $z = 5$.

To achieve the desired resolution, we choose the size of the computational box to be $L = 3.2\ Mpc$ (comoving). Our calculation thus has a top (sub) grid spatial resolution of 6.25 $kpc$ (1.56 $kpc$) and baryonic mass resolution of $8.5 \times 10^4\ M_\odot$ ($1.3 \times 10^3\ M_\odot$) at redshift 3. The results presented in this paper are from a statistical analysis of the top grid only. Results from the higher resolution sub-grid calculation are more relevant for detailed morphology and confinement studies of individual clouds and will be discussed in Zhang, Anninos & Norman (1995).

## 2. RESULTS

We employed the same simple criterion used by Cen et al. (1994) to identify Ly$\alpha$ clouds in our simulations as those structures associated with baryonic overdensity ($\rho/\bar{\rho}$) larger than unity. Although we have extracted data for the smaller density clouds $N_{HI} < 10^{13}\ cm^{-2}$ which are typically associated with underdense regions or voids (Zhang, Anninos & Norman 1995), we consider only the overdense regions in this present work, so a direct comparison of our results can be made to those of Cen et al. (1994). We discuss the effect of the smaller density clouds where appropriate. The remainder of this section summarizes our results and comparisons to observed data.

We find a morphological distinction between high and low column density clouds as shown clearly by Fig. 1, where we plot contours of the projected (over a distance of 0.8 $Mpc$, comoving) column densities for $HI$ and $HeII$. Three contour levels are plotted for each species: $10^{15}$, $10^{17}$ and $10^{19}\ cm^{-2}$. The high column density clouds are typically isolated structures associated with dark matter halos and are found at the intersection of several large filaments. Intermediate and low density absorptions are mainly associated with the filaments or sheets themselves. Notice that each $HI$ contour level is encircled by the equivalent $HeII$ contour. If the $N_{HI} = 10^{14}\ cm^{-2}$ contour is plotted, it will overlap roughly with the $N_{HeII} = 10^{15}\ cm^{-2}$, suggesting that the column density ratio $N_{HeII}/N_{HI}$ approximately equals ten. This is consistent with the spectral shape of the ionizing radiation that we have input. It is of interest to note that a significant number of absorption lines arise from the empty regions or voids with $\rho/\bar{\rho}$ between 0.1 and unity with corresponding column densities in the range $10^{11}$ to $10^{14}\ cm^{-2}$. In fact, we find that most of the $10^{14}\ cm^{-2}$ clouds are associated with overdense regions, while most absorption lines at $10^{11}\ cm^{-2}$ come from the underdense regions. We will explore this important issue in a follow-up paper (Zhang, Anninos, & Norman 1995).

The neutral hydrogen column density distribution is usually believed to be a power law

$f(N_{HI}) \propto N_{HI}^{-\beta}$ with $\beta$ between 1.4 and 1.7 (Carswell 1989) with a possible steepening at column densities larger than $10^{15} \sim 10^{16}$ $cm^{-2}$ (Petitjean et al. 1993; Meiksin & Madau 1993; Hu et al. 1995). The fit to our numerical data at $z = 3$ (shown in Fig. 2a along with the observed data from Petitjean et al. 1993) gives a power law with index of $\beta = 1.55$ with uncertainties of $\pm 0.016$. We also find an apparent deficiency of clouds for column densities larger than $10^{15}$ $cm^{-2}$. This deficiency may be due to the transition from pressure to gravitational confinement of the clouds (Charlton et al. 1994). The integrated column density distribution $f(N > N_{HI})$ is plotted in Fig. 2b as in Cen et al. (1994). A comparison to the data of Rauch et al. (1991), Carswell et al. (1991) and Sargent et al. (1989) shows excellent agreement. Fig. 2a also shows the column density distribution for $HeII$. The results are fit nicely by a power law as for the $HI$ case, with an exponent of $\beta = 1.54$. The data also exhibits a steepening at column densities larger than $10^{16}$ $cm^{-2}$. We note that the turnover at small column densities in Fig. 2b is due to the neglect of underdense regions in the sampling of Ly$\alpha$ clouds and thus the lower density clouds ($< 7 \times 10^{13}$ $cm^{-2}$) are not included in fitting the power law exponent in Fig. 2a. In fact, when the density cutoff is lowered to $\rho/\overline{\rho} = 0.1$, we find no discernible flattening down to $\sim 10^{11}$ $cm^{-2}$, consistent with recent Keck data (Hu et al. 1995).

We have computed the Doppler parameter $b = \sqrt{2k_B T/m_H + v_p^2}$ directly according to the physical state of the clouds by averaging the temperature and the projected line-of-sight velocity throughout each individual structure. This approach is just an approximation since the $HI$ and $HeII$ absorbing regions are generally colder than the entire overdense region selected out through $\rho/\overline{\rho} > 1$, so our estimates of the Doppler parameter are likely to be somewhat larger than the true values. A better but more complicated procedure in computing the Doppler parameter is to generate spectra and fit the resulting profiles. We are currently extending our analysis to this approach as it is more directly comparable to the procedure used by observers (Zhang et al. 1995). The raw data for $z = 3$ (shown in Fig. 3a), which peaks at 27.8 $km\ s^{-1}$, is fit to a Gaussian profile with a median of $<b> = 35.6\ km\ s^{-1}$ and dispersion $\sigma = 14.5\ km\ s^{-1}$ which agrees very well with the observed values $<b> = 36\ km\ s^{-1}$ and $\sigma = 18\ km\ s^{-1}$ of Carswell (1989) but higher than the new Keck observation of $<b> = 28\ km\ s^{-1}$ and $\sigma = 10\ km\ s^{-1}$ (Hu et al. 1995) The distribution of $b$ shows a non-Gaussian tail towards high column densities which is also observed (Carswell 1989) and can be attributed to the non-thermal bulk motion inside the clouds. The median $HI$ Doppler parameter calculated from considering only thermal motion is 15.2 $km\ s^{-1}$, corresponding to an average temperature of $1.4 \times 10^4\ K$. This suggests that both thermal and turbulent motion inside the clouds play equally important roles in the broadening of lines. The Doppler parameter for $HeII$ (Fig. 3b) has a similar Gaussian distribution with a median of $<b> = 31.5\ km\ s^{-1}$ and dispersion $\sigma = 15.5\ km\ s^{-1}$. The ratio of Doppler parameters $b_{HeII}/b_{HI}$ should lie in between the values for a pure thermal broadening (0.5) and the value for a pure bulk velocity broadening (1.0). Our numerical results give an averaged value of 0.87 over all column densities, which matches the observational value of 0.8 (Songaila et al.'s (1995) interpretation of the data provided by Cowie et al. (1995)) extremely well.

The correlation (or lack thereof) between the Doppler parameter and the column density has important implications for the physical state and confinement of the clouds. Pettini et al. (1990) claimed to have found a strong correlation which was subsequently fitted by Bajtlik & Duncan (1991) as $b = 14(\log N_{HI} - 171)$. However this has not been confirmed by other observers, and more recent Keck observations (Hu et al. 1995) do not show any correlation. In Figs. 4 we show scatter plots between the Doppler parameter and the column density. Fig. 4a is constructed assuming only thermal broadening. In this case we observe an apparent correlation which we fit as $b = 6.5(\log N_{HI} - 68.6)$. We compare this result to Fig. 4b where the Doppler parameter is derived from both thermal and bulk velocity broadening. The apparent correlation observed in Fig. 4a is not present here.

Fig. 4b also clearly shows a lower cutoff $b_c$ that is approximately 14 $km\ s^{-1}$ at column density $N_{HI} \sim 10^{13}\ cm^{-2}$, increasing slightly to 21 $km\ s^{-1}$ at $10^{15}\ cm^{-2}$ in a roughly linear fashion as $b_c = 3.5 \log N_{HI} - 32$. For column densities near $10^{14}\ cm^{-2}$, we can expand the logarithmic term to rewrite the fit as $b_c = 3.5 N_{HI}/10^{14} + 13.5$, which is in excellent agreement with the lower envelope $b = 4 N_{HI}/10^{14} + 16$ found in recent Keck observations (Hu et al. 1995).

The results presented thus far have been for a fixed redshift $z = 3$. However, any viable model for the Ly$\alpha$ forest must also predict the evolution of cloud properties. The number of Lyman forest clouds per unit redshift has been observed to depend rather strongly on redshift, suggesting that these clouds are rapidly dissipating. This evolution is usually fit to a power law $dN_c/dz \propto (1+z)^\gamma$ with $\gamma = 2.3 \pm 0.4$ as the observed value for $N_{HI} \geq 10^{14}\ cm^{-2}$ (Murdoch et al. 1986; Bajtlik et al. 1988; Lu et al. 1991). The exact value of $\gamma$ in numerical simulations will reflect both changes in the ionizing radiation and the merging of structures. We have calculated the evolution index $\gamma$ for clouds with column densities $10^{14} \leq N_{HI} \leq 3 \times 10^{17}\ cm^{-2}$ from redshift 3 to 4.2. We find $\gamma = 2.22$, in excellent agreement with observations. We also find $\gamma = 3.89$ for clouds with column densities $10^{15} \leq N_{HI} \leq 3 \times 10^{17}\ cm^{-2}$, and $\gamma = 4.21$ for clouds with column densities $10^{16} \leq N_{HI} \leq 3 \times 10^{17}\ cm^{-2}$. This trend of stronger evolution towards higher column densities is consistent with the observations of Bechtold (1994), although we note that this issue is not settled observationally (Giallongo 1991).

## 3. CONCLUSIONS

We find that our cosmological model with microphysical processes are consistent with all of the observational results to which we have made comparisons. This includes $HI$ column density distributions, Doppler parameter distributions, and cloud number evolution. We also find that the geometries of the clouds are linked to the local thermal and gravitational environments. Spherical-shaped clouds are concentrated at the intersection of filaments and are typically more massive than the sheet-like clouds which tend to lie along the filament strands.

We have also computed various properties of $HeII$ Ly$\alpha$ absorption lines which are now becoming accessible to observers. The ratio of Doppler parameters $b_{HeII}/b_{HI}$ is consistent with observed values. However, for the most part, our $HeII$ results can be regarded as predictions of the standard CDM model and the specific microphysical processes that we have accounted for in the present model. Further comparisons of Helium lines may have to await new high resolution UV satellite data.

This work is currently being extended to higher grid resolution which can confirm these results and be used to investigate the morphology issues in greater detail (Zhang, Anninos & Norman 1995). A current limitation of our simulations is the lack of large scale power due to the finite size of the numerical grid. Larger grid simulations are underway to investigate the effect of neglecting long wavelength perturbations. Finally we note that to accurately model high column density clouds with $N_{HI} > 3 \times 10^{17}\ cm^{-2}$ (Lyman limit and damped Ly$\alpha$), it is important to include radiation transfer since those clouds can shield themselves from the ionizing UV flux.

We thank Renyue Cen, Art Davidsen, Limin Lu, Avery Meiksin, Paul Shapiro, Michael Rauch, David Weinberg, and especially our referee Jane Charlton for many enlightening discussions. We also thank Tom Abel for providing us with his fits to several rate coefficients prior to publication and for many conversations regarding the chemical model. This work is done under the auspices of the Grand Challenge Cosmology Consortium (GC3) and supported in part by NSF grant ASC-9318185. The simulations were performed on the CONVEX-3880 at the National Center for Supercomputing Applications, University of Illinois at Urbana-Champaign.

---

Fig. 1.— Contour plots of $HI$ and $HeII$ column densities at redshift 3: The size of the slice (1.6 $Mpc$ comoving) is half of our computational box size, and the projection length (0.8 $Mpc$ comoving) is one quarter of our box size. The solid lines are contour levels for $HI$ and the dotted lines are for $HeII$. Three contour levels corresponding to column densities of $10^{15}$, $10^{17}$ and $10^{19}$ $cm^{-2}$ are shown.

Fig. 2.— Column density distribution per unit redshift at $z = 3$. (a) The distribution per unit column density: The open circles are our numerical data for $HI$, the solid line is a power-law fit with exponent $\beta = 1.55$, and the crosses are observed data from Petitjean et al. (1993). Note we converted their data to per unit redshift instead of per unit absorption length. The filled circles are our numerical data for $HeII$ and the dashed line is a power-law fit with exponent $\beta = 1.54$. We note that there is an apparent deficiency of clouds at column densities greater than $10^{15}$ and $10^{16}$ for $HI$ and $HeII$ respectively. (b) The integrated distribution: The solid line is for $HI$. The filled circles are observational data from Rauch et al. (1992), the open circle is from Carswell et al. (1991) and the filled square is from Sargent et al. (1989). Also shown is the column density distribution for $HeII$ (dashed line).

Fig. 3.— Distribution of Doppler parameter at $z = 3$. (a) $HI$: The dotted line is a Gaussian fit to our numerical data with median $< b >= 35.6$ $km$ $s^{-1}$ and dispersion $\sigma = 14.5$ $km$ $s^{-1}$. (b) $HeII$: The dotted line is a Gaussian fit to our numerical data with median $< b >= 31.5$ $km$ $s^{-1}$ and dispersion $\sigma = 15.5$ $km$ $s^{-1}$. Note that both $HI$ and $HeII$ distributions exhibit a broad wing towards large values of $b$.

Fig. 4.— A comparison of scatter plots for the $HI$ Doppler parameter $b$ versus column density. (a) Thermal broadening only: The dashed line is a fit by Bajtlik & Duncan (1991) for the observational data of Pettini et al.(1990). Our data also suggests a correlation which we fit as $b = 6.5(\log N_{HI} - 68.6)$, shown by the solid line. (b) Thermal plus bulk velocity broadening: The apparent correlation observed in (a) is not present here. There is a noticeable lower cutoff in $b$ that increases linearly with column density as $b_c = 3.5 \log N_{HI} - 32$ (solid line), and is consistent with that found by Hu et al (1995) for column densities $\lesssim 10^{14}\ cm^{-2}$.

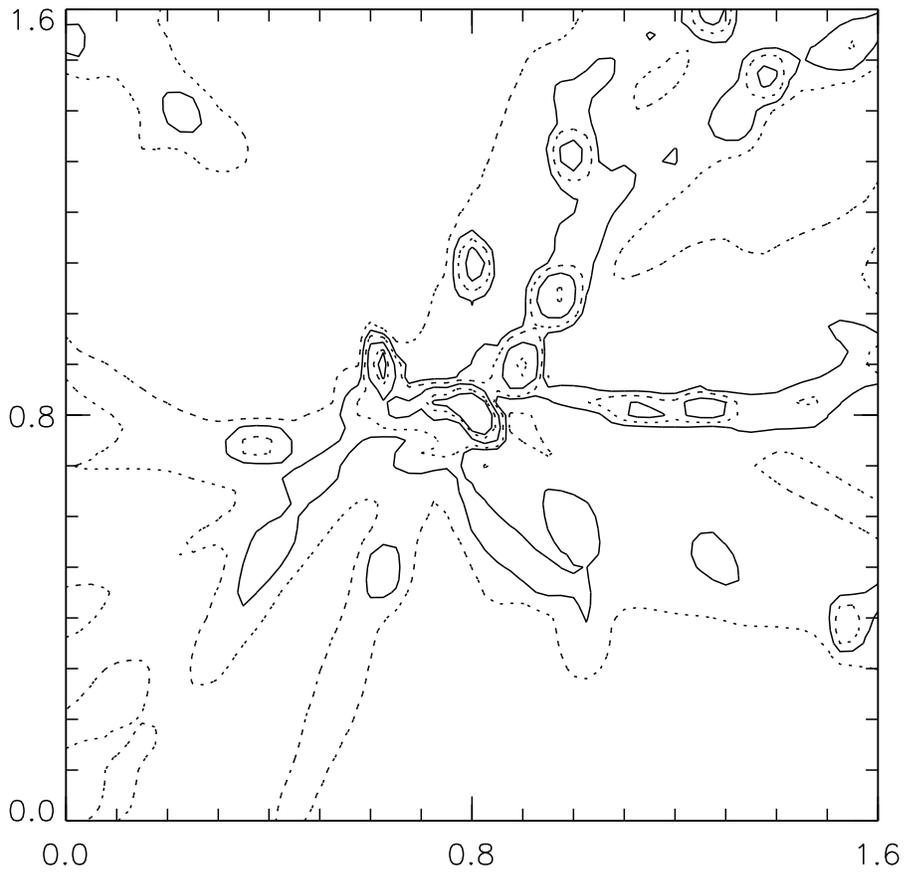

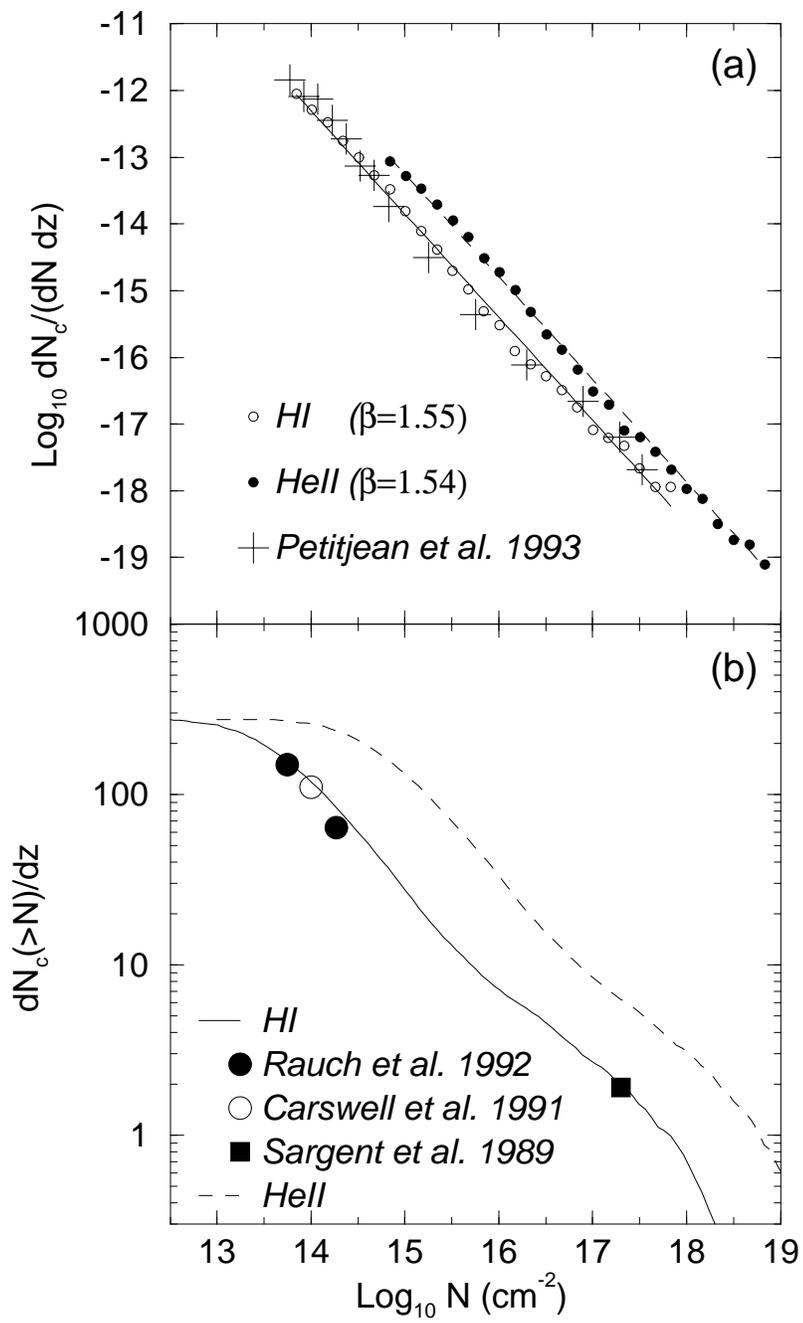

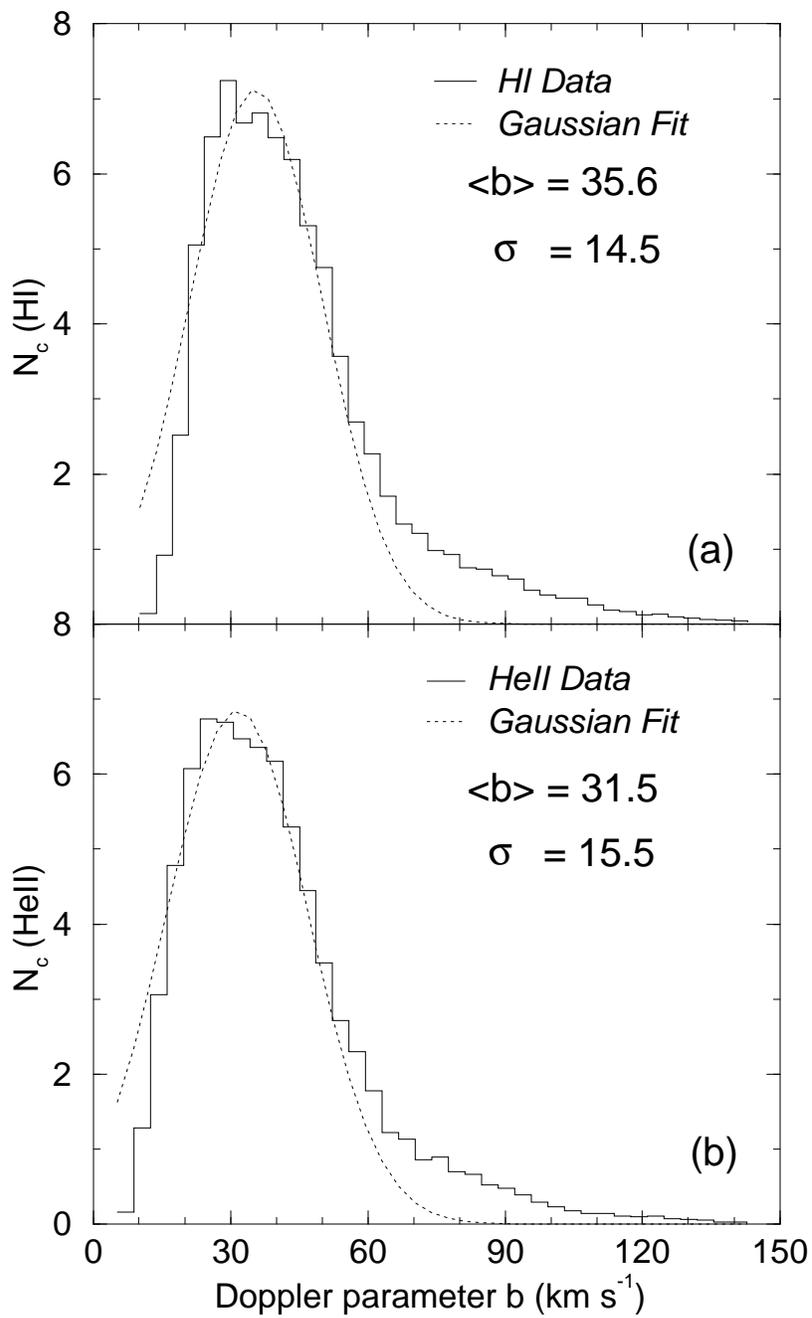

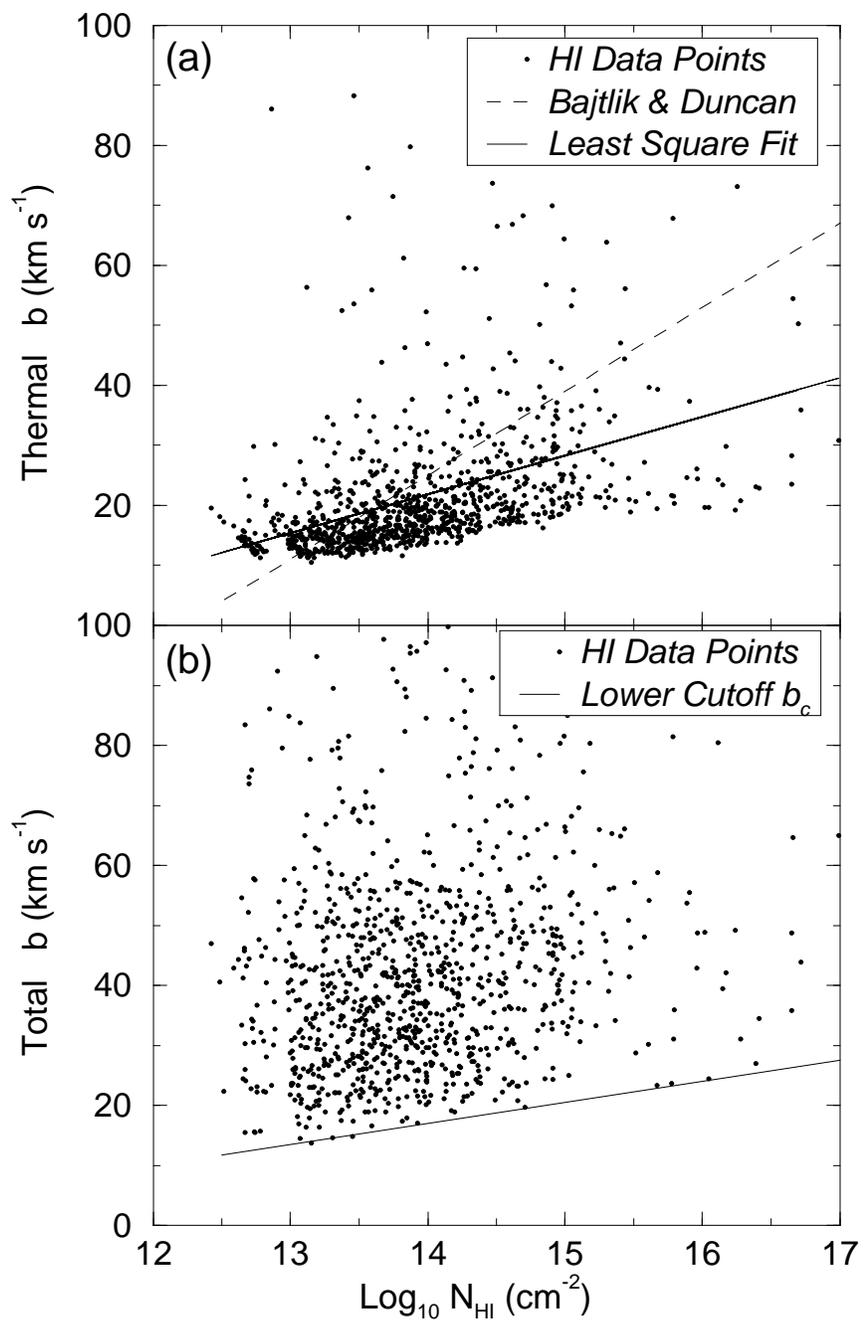